\documentclass[final,letter]{elsart} 
\usepackage[super, sort]{natbib}
\bibpunct{(}{)}{;}{a}{,}{,} 
\usepackage{amsmath,amssymb}
\usepackage{graphicx}
\usepackage{lineno}
\usepackage{setspace}
\usepackage{rotating}

\def\tableline{ \vskip .1in \hrule height.6pt \vskip 0.1in}
\newcommand{\ppr}{$\lambda_1(\mathbf{A})$ }
\DeclareMathOperator{\diag}{diag}

\begin{document}

\begin{frontmatter}
\title{Functional Integration of Ecological Networks through Pathway Proliferation}
\author[TIE,skio]{Stuart R. Borrett\corauthref{cor1}},
\ead{sborrett@stanford.edu}
\author[TU]{Brian D. Fath}, and
\author[TIE]{Bernard C. Patten}


\corauth[cor1]{Current Address: Computational
Learning Laboratory, Center for the Study of Language and
Information, Stanford University, Stanford, CA, 94305 USA}
\address[TIE]{Institute of Ecology, University of Georgia, Athens, GA, 30606 USA}
\address[skio]{Skidaway Institute of Oceanography, Savannah, GA 31411 USA}
\address[TU]{Biology Department, Towson University, Towson, MD 21252 USA }

\begin{abstract}
Large-scale structural patterns commonly occur in network models of complex systems including a skewed node degree distribution and small-world topology.  These patterns suggest common organizational constraints and similar functional consequences.  Here, we investigate a structural pattern termed pathway proliferation. Previous research enumerating pathways that link species determined that as pathway length increases, the number of pathways tends to increase without bound. We hypothesize that this pathway proliferation influences the flow of energy, matter, and information in ecosystems.  In this paper, we clarify the pathway proliferation concept, introduce a measure of the node--node proliferation rate, describe factors influencing the rate, and characterize it in 17 large empirical food-webs.  During this investigation, we uncovered a modular organization within these systems.  Over half of the food-webs were composed of one or more subgroups that were strongly connected internally, but weakly connected to the rest of the system.  Further, these modules had distinct proliferation rates.  We conclude that pathway proliferation in ecological networks reveals subgroups of species that will be functionally integrated through cyclic indirect effects.
\end{abstract}

\end{frontmatter}

\begin{spacing}{1}
\section{Introduction and Motivation}
Large-scale structural patterns have been uncovered in network models of complex systems, suggesting the possibility of common organizational constraints and similar functional consequences. Network models are mathematical graphs composed of nodes and undirected edges or directed arcs that connect the nodes.  For example, in a social network nodes might represent individuals in a community and the edges or links could represent a social relationship between the individuals such as collaboration \citep{newman01,newman01phyrev}.  In a model of the World Wide Web, web pages are nodes connected by hyperlinks \citep{albert99,barabasi99}.  Ecologists use network models in many ways, including to represent trophic relations in food-webs and more generally energy--matter flux in ecosystems \citep{cohen90,higashi91,margalef63,pimm82}.  In these networks, species or functional groups form the node set while the presence of energy and matter transfers and transformations are represented by links.

Traditionally, complex systems have been modeled using random graphs \citep{erdos59,erdos60,gardner70,may72}.  However, ecologists have demonstrated that random graphs are inadequate models of ecological systems; food-web and ecosystem models often contain structures not commonly found in random graphs \citep{cohen90,deangelis75,pimm79,pimm82,pimm91,ulanowicz91,lawler78}.  In food-webs, these hypothesized structures include short food-chain lengths \citep{pimm77,post02} and little or no cycling \citep{cohen90}.  In addition, several forms of modularity---hierarchic compartmentalization into subsystems---have been hypothesized for food-webs and ecosystems \citep{allen82,may72,pimm79,pimm80,yodzis82}.  Furthermore, \citet{ulanowicz91} demonstrated that random networks (based on the Poisson, uniform, Gaussian, negative exponential, and log-normal probability distribution functions) failed to capture the distribution of connections in real ecosystems.  Ecologists hypothesize that these structural differences exist because ecological systems are shaped and constrained by thermodynamic laws and natural history \citep{jorgensen92,lawler78,muller98,williams00}.

Likewise, investigations of other types of complex systems have identified a number of distinctive patterns common in complex systems not found in purely random graphs \citep{albert02,newman02,newman03,moreno38,price65}.  For example, the distribution of node degree (i.e., the number of edges (links) incident to a node) is often skewed in models of complex systems, following an exponential distribution or a power-law distribution rather than the Poisson distribution of random graphs.  The power-law distribution was found in the World Wide Web \citep{barabasi99}, metabolic networks \citep{jeong00}, and some but not all food-webs \citep{dunne02el, montoya02}.  The power-law degree distribution implies that there are a large number of nodes with very few connections, while a few nodes have a large number of connections \citep{barabasi02}.  This topology tends to increase network robustness to random node or edge deletion, while making it more sensitive to targeted attacks \citep{albert00,dunne02el}.  The small-world pattern is another commonly found topology \citep{watts99,watts98}.  In small-world networks, the degree of node clustering is larger and the maximum distance (where distance is the shortest path between two nodes) is lower than expected from random graphs.  This arrangement tends to increase the transmission speed of diseases, energy, matter, and information through networks.  The largest distance in food-web graphs tends to be small, but the degree of clustering varies \citep{dunne02pnas}.

Pathway proliferation is another large-scale topological characteristic of networks, with implications for energy, matter, and information transmission.  It is the tendency for the number of pathways in a network to increase geometrically without bound as pathway length increases.  Patten and colleagues \citep{patten85a,patten85b,patten82} first observed this tendency in small, well-connected ecosystem models during the early development of ecosystem network analysis.  More recently, \citet{fath98diss} and \citet{borrett03} showed that the rate of pathway proliferation is variable among networks. This is significant because the pathway proliferation rate characterizes how quickly the number of indirect pathways increases, and thus, the number of pathways available for interactions.  Food-web investigations often emphasize the shortest pathways, assuming that most significant interactions occur over these routes \citep[e.g.,][]{caldarelli98,post00}.  However, previous results from Network Environ Analysis, an environmental application and extension of economic Input--Output Analysis, indicate that flows over longer indirect pathways can be significant or even dominant constituents of total system throughflow \citep{higashi86,higashi89,patten83}, which is a measure of whole system activity.  This result has important implications for trophodynamics \citep{burns91,patten90,whipple98} and biogeochemical cycling in ecosystems \citep{borrett06,finn80,patten76}.  Given the possible significance of indirect pathways in network models of conservative transport systems like ecosystems, it is critical to understand the network characteristics influencing the pathway proliferation rate.

In this paper, we clarify the pathway proliferation concept, describe factors influencing the proliferation rate, and characterize pathway proliferation rates in 17 large empirical food-webs.  In Section \ref{sec:pp} we review relevant mathematics to build a better understanding of pathway proliferation.  In Section \ref{sec:fw} we apply this understanding to 17 food-web models drawn from the literature.  This analysis reveals a type of modularity in some of the food-webs, lending support to the hypothesis that food-webs have a modular structure \citep{krause03,may72,pimm80,yodzis82,allesina2005}.  We conclude by summarizing our findings, discussing their relevance for ecological systems, and suggesting next steps along this research path.
\section{Pathway Proliferation}\label{sec:pp}
Although Patten and colleagues \citep{patten85a,patten85b,patten82} introduced pathway proliferation into the ecological literature over two decades ago, it is not well understood.  In this section, we synthesize mathematical results from graph theory and matrix algebra to determine a method for quantifying the node--node pathway proliferation rate, and to identify the bounds and expected value of the rate.  In addition, we uncover the possibility of differing rates of pathway proliferation for modules within a network.

Network models of complex systems are mathematically graphs which can be directed or weighted \citep{bangjensen01,ponstein66}.  A graph $G$ is specified by a set of $n$ nodes and $e$ unoriented edges ($0 \le e \le n(n-1)/2 + n$), where edges indicate an undirected relationship between two nodes. A directed graph (digraph) $D$ is also specified by a set of $n$ nodes, but instead of edges it has $L$ oriented arcs or links ($0 \le L \le n^2$).  Digraph structure is partially characterized by two connectivity measures, connectance $C = L/n^2$ and link density $L/n$, which are common metrics in the food-web literature \citep{cohen90,martinez91}.  Edges and links can be assigned weights to represent the relationship strength.

In this paper, we focus on simple unweighted digraphs, where simple implies no more than one link from one node to any other.  We do this for two reasons.  First, directed graphs are often appropriate for ecological applications as many ecological processes are oriented (e.g., predation and excretion generate energy and matter flows from one ecosystem element to another).  Second, while the mathematics described in this paper may apply to non-simple and weighted graphs, our interest here is primarily network structure as it is a necessary element of understanding ecosystem organization.
\subsection{Quantifying Pathway Proliferation}
Here, we review relevant definitions and results from graph theory and linear algebra.  We first show why the dominant eigenvalue of a strongly connected digraph is a good measure of the node--node proliferation rate.  We conclude by describing how the dominant eigenvalue can distinguish between fundamentally different ecological networks.

\begin{figure}[t]
\begin{center}
\includegraphics[scale=0.6]{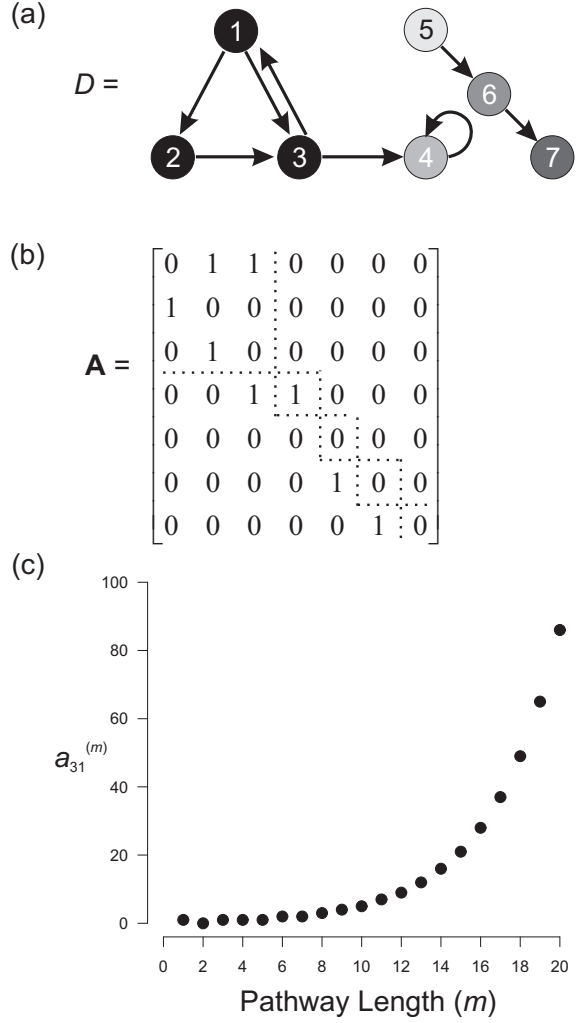}
\caption{Example digraph, its associated adjacency matrix, and pathway proliferation. (a) Digraph $D$ composed of strongly connected components $\mathbf{K}_1 = \{1,2,3\}$, $\mathbf{K}_2 = \{4\}$,  $\mathbf{K}_3 = \{5\}$, $\mathbf{K}_4 = \{6\}$, $\mathbf{K}_5 = \{7\}$, (b) the adjacency matrix $\mathbf{A}$ associated with the digraph $D$ (dotted lines denote strongly connected components), and (c) a plot demonstrating pathway proliferation from node 3 to 1: the number of pathways $a_{31}^{(m)}$ increases as pathway length $m$ increases.  Transient effects created by smaller eigenvalues are visible in the first ten pathway lengths, but the rate of pathway proliferation has nearly converged to $\lambda_1(\mathbf{K}_1) = 1.32$ by a pathway length of 20.}
\label{fig1}
\end{center}
\end{figure}

In a directed graph, a pathway is an alternating sequence of nodes and links connecting a starting and a terminal node.  Pathway length $m$ is the number of links in the pathway.  For example, in the directed graph $D$ in Figure~\ref{fig1}a there is a pathway of length 2 from node 1 to node 3 (e.g., 1 $\rightarrow$ 2 $\rightarrow$ 3).  Cycles are pathways with the same starting and terminal nodes, and a cycle of length one is a self-loop.  In our example network, 1 $\rightarrow$ 2 $\rightarrow$ 3 $\rightarrow$ 1 is cycle of length three, and 4 $\rightarrow$ 4 is a self-loop.  Pathways with self-loops are termed walks, those without are paths (Patten, 1985a).  Alternatively, we can represent $D$ with its associated and isomorphic adjacency matrix $\mathbf{A}_{n\times n} = (a_{ij})$, where $a_{ij} = 1$ if and only if there is a link from $j$ to $i$ (note column to row orientation), otherwise $a_{ij} = 0$ (Figure~\ref{fig1}b). The number of direct links terminating or starting at a node is termed the in- and out-degree, respectively.  These are calculated as $\mathbf{k}^{in} = (k^{in}_i) = \sum^n_{j=1}a_{ij}$  and $\mathbf{k}^{out} = (k^{out}_j) = \sum^n_{i=1}a_{ij}$, where $\mathbf{k}^{in}$ and $\mathbf{k}^{out}$ are $1 \times n$ and $n \times 1$ vectors respectively.  Average in- and out-degrees ($<k^{in}>$ or $<k^{out}>$), and degree distributions $P(k)$ are ways of characterizing network structure \citep{albert02,newman03,ulanowicz91}.

Indirect pathways $(m>1)$ are enumerated by raising the adjacency matrix to the $m^{th}$ power $\mathbf{A}^m = (a_{ij})^m$ \citep{ponstein66}.   Again, pathway proliferation is the tendency for the number of pathways in a network to increase without bound as a function of increasing pathway length (Figure~\ref{fig1}c).  \citet{borrett03} approximated whole-system pathway proliferation rate as
\begin{displaymath}
\frac{log(\sum\sum a_{ij}^{(m)})}{log(\sum\sum a_{ij}^{(m-1)}))}
\end{displaymath}
where $a_{ij}^{(m)}$ signifies the $a_{ij}$ element of $\mathbf{A}^m$, but this aggregate estimate only holds for sufficiently large $m$.  Therefore, following \citet{fath98diss}, here we will define pathway proliferation in terms the rate at which $a_{ij}^{(m)}$ changes as $m \rightarrow \infty$.  This makes pathway proliferation a combinatorial property of $\mathbf{A}$ \citep{seneta73}.

As $m$ increases, the development of $\mathbf{A}^m$ is determined by its characteristic polynomial, $\pi(\lambda)=\det(\lambda \mathbf{I} - \mathbf{A})$, where $\lambda$ is the variable in the polynomial and $I_{n\times n}$ is the identity matrix \citep{godsil93,seneta73}.  The eigenvalues ($\lambda_i$,  $i = 1,\ldots,n$) of $\mathbf{A}$ are roots of the characteristic polynomial, determined as solutions to $\pi(\lambda)=0$.  The set of eigenvalues $\{\lambda_1 \ge \lambda_2 \ge \ldots \ge \lambda_i \ge \ldots \ge \lambda_n\}$ is the spectrum of $\mathbf{A}$, which \citet{cvetkoviac80} study to determine structural properties of graphs.  In undirected graphs, $\mathbf{A}$ is symmetric and all $\lambda_i$ are real; in directed graphs $\lambda_i$ may be a complex number.

In addition to being the roots of the characteristic polynomial, the eigenvalues must satisfy $\mathbf{AW=\Lambda W}$, where $\mathbf{W}_{n\times n} = \left[ \begin{array}{c|c|c|c|c|c} \mathbf{w}_1 & \mathbf{w}_2 & \ldots & \mathbf{w}_i & \ldots & \mathbf{w}_n \end{array} \right]$ is a composite matrix of the right eigenvectors $\mathbf{w}_i$, and $\mathbf{\Lambda}_{n\times n} = \diag(\lambda_i)$ is a matrix with the eigenvalues of $\mathbf{A}$ on the principle diagonal and zeros in all other positions.  Further, the eigenvalues must satisfy $\mathbf{VA=V\Lambda}$, where $\mathbf{V}_{n\times n} = \left[ \begin{array}{c|c|c|c|c|c}
\mathbf{v}_1 & \mathbf{v}_2 & \ldots & \mathbf{v}_i & \ldots & \mathbf{v}_n \end{array} \right]$ is a composite matrix of the left eigenvectors $\mathbf{v}_i$. $\mathbf{A}^m$ is then determined as
\begin{eqnarray}\label{eq:eig}
\mathbf{A}^m &=& \mathbf{W}\times\mathbf{\Lambda}^m\times \mathbf{W}^{-1}, \\ \nonumber
\mathbf{A}^m &=& \mathbf{W}\times\left[ \begin{array}{cccc}
                 \lambda_1^m & 0 & \cdots & 0\\
                 0 & \lambda_2^m & \cdots & 0\\
                 \vdots & \vdots & \ddots & \vdots\\
                 0 & 0 & \cdots & \lambda_n^m\\
                 \end{array} \right]\times \mathbf{W}^{-1}
\end{eqnarray}
where $\mathbf{W}$ is as before and $\mathbf{W}^{-1}$ is the matrix inverse of $\mathbf{W}$, which are the transposed complex conjugates of the left eigenvectors $\mathbf{V}$ \citep{caswell01}.  If we let $\mathbf{v}_i^\star$ be the i$^{th}$ row of $\mathbf{W}^{-1}$, then we can rewrite equation \ref{eq:eig} as
\begin{equation}\label{eq:Am}
\mathbf{A}^m = \lambda_1^m\mathbf{w}_1\mathbf{v}_1^\star + \lambda_2^m\mathbf{w}_2\mathbf{v}_2^\star+ \cdots +\lambda_i^m\mathbf{w}_i\mathbf{v}_i^\star +\lambda_n^m\mathbf{w}_n\mathbf{v}_n^\star
\end{equation} 
Equation \ref{eq:Am} is the pathway generating function of $\mathbf{A}$ \citep{godsil93}, and illustrates how the development of $\mathbf{A}^m$ depends on the spectrum of $\mathbf{A}$.

Next, we introduce two classification schemes, one for digraphs and one for matrices, because they are required to apply a theorem that will let us develop a succinct estimate of the pathway proliferation rate.  In the first scheme, digraphs are classified as one of three types: strongly connected, weakly connected, and disconnected.  A digraph is strongly connected (strong) if it is possible to reach every node from every other over a pathway of unspecified length.  \citet{bangjensen01} define a (sub)digraph with only one node as strong, though this is trivial for our purposes.  A weakly connected (weak) digraph is one in which it is possible to reach any node from any other node if we ignore link orientation, but it is impossible when following link orientation.  Nodes of a weak digraph must have an in-degree or out-degree of at least 1.  A disconnected graph is one that contains one or more non-adjacent strong or weak components.  The digraph in Figure~\ref{fig1}a is disconnected.

The second classification scheme distinguishes three types of adjacency matrices.  A matrix is irreducible if and only if it is associated with a strong digraph, while one associated with a weakly connected or disconnected digraph is reducible \citep{berman79}.  Irreducible matrices are further divided into two classes: primitive and imprimitive.  A primitive matrix is an irreducible matrix that becomes positive ($a_{ij}>0$, for all $i,j$) when raised to a sufficiently large power \citep{seneta73}.  Furthermore, all adjacency matrices are non-negative because all $a_{ij}$ are greater than or equal to zero.

Weak and disconnected digraphs are decomposable into a unique set of maximally-induced, strong subdigraphs ($\mathbf{K}_i$, $i=1,\ldots,\alpha$, where $\alpha\le n$), that are termed \textit{strongly connected components} \citep{bangjensen01}.  An induced subdigraph of $D$ is a subset of nodes in $D$ with all links that both start and terminate on the node subset, and a maximally induced subdigraph is the largest one that is strong.  This implies that there is at least one simple cycle (no repeated medial nodes) that connects all nodes in a non-trivial strongly connected component.  Further, adjacency matrices associated with $\mathbf{K}_i$ are irreducible.  For example, the digraph in Figure~\ref{fig1}a contains two connected subdigraphs $\{1, 2, 3, 4\}$ and $\{5, 6, 7\}$.  Furthermore, it can be partitioned into five strongly connected components $\mathbf{K}_1 = \{1, 2, 3\}$, $\mathbf{K}_2 = \{4\}$, $\mathbf{K}_3 = \{5\}$, $\mathbf{K}_4 = \{6\}$, $\mathbf{K}_5 = \{7\}$, of which only $\mathbf{K}_1$ is non-trivial.  The adjacency matrix associated with each strongly connected component is irreducible; the adjacency matrices associated with $\mathbf{K}_1$ and $\mathbf{K}_2$ are primitive.

Given these definitions, the Perron--Frobenius theorem guarantees there is one real eigenvalue equal or larger than all other eigenvalues, $\lambda_1\ge\lambda_i$ ($i = 2,\ldots,n$) in irreducible matrices \citep{berman79,seneta73}.  In the literature \ppr is alternately referred to as the dominant eigenvalue, the Perron eigenvalue, and the spectral radius.  Next, we illustrate why \ppr is a good measure of the pathway proliferation rate.

As shown by \citet{caswell01}, we can divide both sides of equation \ref{eq:Am} by $\lambda_1$  to obtain,
\begin{equation}\label{eq:rate}
\frac{\mathbf{A}^m}{\lambda_1^m} = \mathbf{w}_1\mathbf{v}_1+ \frac{\lambda^m_2}{\lambda_1^m}\mathbf{w}_2\mathbf{v}_2+\frac{\lambda^m_3}{\lambda_1^m}\mathbf{w}_3\mathbf{v}_3+ \cdots+\frac{\lambda^m_n}{\lambda_1^m}\mathbf{w}_n\mathbf{v}_n.
\end{equation}
If $\mathbf{A}$ is primitive and irreducible, then $\lambda_1$ is strictly larger than $\|\lambda_i\|$ for all $i>1$, where $\|\bullet\|$ is the norm of $\bullet$  (this is necessary since  $\lambda_i$ may be complex).  Taking the limit of both sides of equation \ref{eq:rate} as pathway length increases, we find that
\begin{equation}
\lim_{m\rightarrow\infty}\frac{\mathbf{A}^m}{\lambda_1^m}=\mathbf{w}_1\mathbf{v}_1.
\end{equation}
Thus, smaller eigenvalues influence pathway proliferation over shorter path lengths (Figure~\ref{fig1}c), but as path length increases the pathway proliferation rate asymptotically becomes $\lambda_1$.  Consequently, $\mathbf{A}^{m+1}/ \mathbf{A}^m\rightarrow\lambda_1$ as $m\rightarrow\infty$ \citep{hill}, and $\lambda_1$ is the asymptotic rate of pathway proliferation in a strongly connected graph with a primitive adjacency matrix.  In addition, the damping ratio
\begin{equation}\label{eq:damp}
\rho = \frac{\lambda_1}{\|\lambda_2\|}
\end{equation}
characterizes the rate of convergence to \ppr \citep{caswell01}.

In strong digraphs with an imprimitive and irreducible adjacency matrix there are $c\le n$ eigenvalues with the same absolute magnitude, and one or more may be complex \citep{seneta73}.  The Perron--Frobenius theorem then indicates that the common absolute magnitude of the $c$ eigenvalues will be larger than the other $n-c$ eigenvalues.  In this case, the dominant eigenvalue has a multiplicity of $c$, and as $m\rightarrow\infty$ only the $c$ largest eigenvalues will influence pathway proliferation.  \citet{caswell01} reports that these digraphs generate oscillatory dynamics.

Given these mathematical results, each strong component $\mathbf{K}_i$ of a weak or disconnected digraph will have an independent rate of pathway proliferation, $\lambda_1(\mathbf{K}_i)$ (read ``$\lambda_1$ of $\mathbf{K}_i$").  The eigenvalues of a reducible matrix are the union set of the eigenvalues of the adjacency matrices associated with strongly connected components \citep{jain03}.  Thus, the maximum dominant eigenvalue of the strongly connected components will be the dominant eigenvalue of the whole digraph.  Further, trivial strongly connected components---those with only one node---will have a pathway proliferation rate of unity or 0 depending on whether or not it has a self-loop.  Thus, if a digraph is composed of only trivial strongly connected components without self-loops, pathway proliferation will not occur.  This is true of all acyclic digraphs, and suggests that we can use the dominant eigenvalue to detect the presence of cycles in digraphs \citep{jain03}.  For this application there are three cases:
\begin{enumerate}
\item if $\lambda_1(\mathbf{A}) = 0$, then $\mathbf{A}$ has no cycles;
\item if $\lambda_1(\mathbf{A}) = 1$, then $\mathbf{A}$ has at least one cycle and all cycles occur in strongly connected components that have only one simple cycle; and
\item if $\lambda_1(\mathbf{A}) > 1$, then $\mathbf{A}$ has more than one simple cycle.
\end{enumerate}
Based on an independent development, \citet{fath98diss} interpreted similar results as three classes of feedback: 1) no feedback, 2) simple feedback, and 3) cyclic feedback in strongly connected networks.  Notice that a graph with $\lambda_1(\mathbf{A})\ge1$ could have a reducible or irreducible adjacency matrix, while the adjacency matrix of a graph with $\lambda_1(\mathbf{A})=0$ is necessarily reducible with $\alpha = n$ trivial strongly connected components.

Similarly, as $m \rightarrow \infty$ we can summarize three possibilities for the dominant eigenvalue as a measure of pathway proliferation in digraphs:
\begin{enumerate}
\item  if $\lambda_1(\mathbf{A}) = 0$, then the number of pathways between two nodes declines to zero;
\item  if $\lambda_1(\mathbf{A}) = 1$, then the number of pathways between nodes in a strongly connected component remains constant; and
\item  if $\lambda_1(\mathbf{A}) > 0$, then the number of pathways between nodes in at least one strongly connected component ($\mathbf{K}_i$) increases without bound at an asymptotic rate equal to $\lambda_1(\mathbf{K}_i)$ where $\max(\lambda_1(\mathbf{K}_i)) = \lambda_1(\mathbf{A})$.
\end{enumerate}
\subsection{Bounds and Expected Values of Pathway Proliferation}
Given that \ppr is the asymptotic rate of pathway proliferation in strong digraphs, it would be useful to know its theoretical bounds and expected value.  Again, existing mathematics provides us with some of these answers.

Matrix theory bounds the dominant eigenvalue of a non-negative matrix by the minimum and maximum column (row) sum, which in the context of directed graphs is the minimum and maximum out-degree (in-degree), where equality holds only if $k^{in}=k^{out}$  \citep{berman79,seneta73}.  Thus, $\max\big(\min(k_j^{in}),\min(k_j^{out})\big)\le\lambda_1(\mathbf{A})\le \min\big(\max(k_j^{in}),\max(k_j^{out})\big)$.  In a strongly connected digraph with more than one node, all nodes must have at least one and a maximum of $n$ incoming and outgoing links.  Therefore, $1\le\lambda_1(\mathbf{A})\le n$ for a strong digraph.  As stated previously, a trivial component with no self-loops will have $\lambda_1(\mathbf{K}_i)= 0$, and a complete graph will have $\lambda_1(\mathbf{A})= n$ (allowing self-loops).  Notice that in the binary matrix $\mathbf{A}$, \ppr cannot take values between 0 and 1.

With these bounds, we can now examine the expected value of $\lambda(\mathbf{A})$.  In undirected random graphs $G$ with $\mathbf{A} = (a_{ij})$ where $a_{ij} = a_{ji} = 1$ with probability $p$ $(0 < p < 1)$ and $a_{ij} = a_{ji} = 0$ with probability $(1-p)$, Juh\'{a}sz proved that
\begin{equation}
\lim_{n\rightarrow\infty}\frac{\lambda_1(\mathbf{A})}{n}=p
\end{equation}
\citep{cvetkoviac90}.  This implies that $\lambda_1(\mathbf{A})\sim np$ in the limit of large $n$.  Furthermore, given \ppr and $n$ we can determine the approximate number of undirected edges in $\mathbf{A}$.  A random graph is not necessarily connected, but \citet{erdos59,erdos60} showed that the fraction of nodes connected in a single component increases rapidly when the average link density exceeds unity.

The expected value of \ppr is sensitive to the assumptions of random graphs.  For example, \citet{farkas01} and \citet{Goh01} found that in random graphs with a power-law distribution of node degrees rather than the standard Poisson distribution, \ppr increased with approximately $n^{1/4}$.  Furthermore, \citet{deaguiar05} demonstrated that the expected \ppr is further modified if the network topology displays a hierarchic modularity.

\begin{figure}[t]
\center
\includegraphics[scale=0.5]{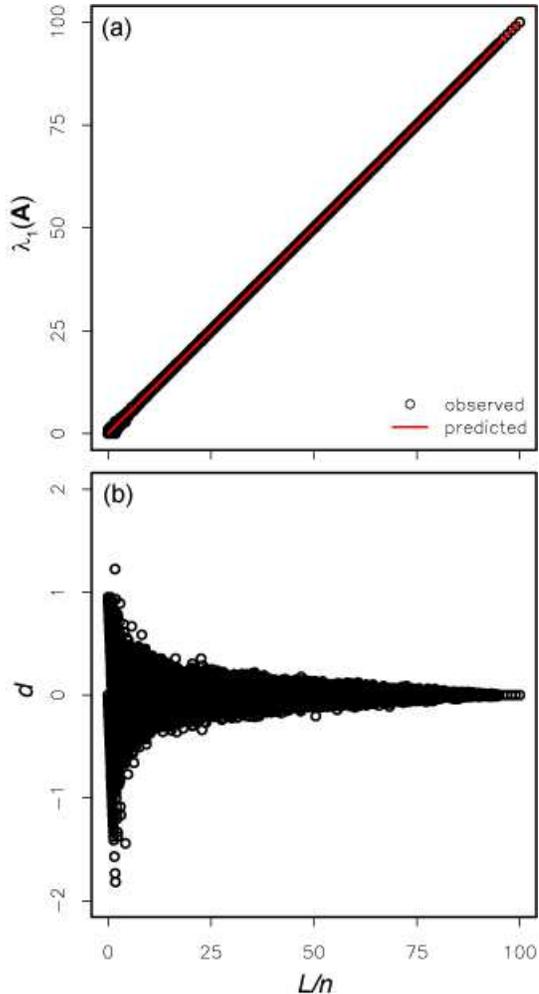}
\caption{Dominant eigenvalue in random digraphs.  (a) Points show the relationship between dominant eigenvalues \ppr and link density $L/n$ in 99,000 uniform random digraphs where $a_{ij} = 1$ with probability $p$ and $a_{ij} = 0$ with probability $(1-p)$ (50 replicates of each combination of $n = \{2,3,\ldots,100\}$ and $p = \{0.05, 0.10,\ldots, 1\}$).  The line indicates the expected $\lambda_1(\mathbf{A})= L/n$ relationship.  (b) Distribution of residual values indicating that as $L/n$ increases it becomes a better predictor of $\lambda(\mathbf{A})$.}
\label{fig:sr}
\end{figure}

Random graphs are well-studied, but properties of random digraphs are less well-known. Some characteristics are similar to undirected graphs.  For example, the in- and out-degree of random digraphs has a Poisson distribution, and when link density is greater than unity the expected size of the largest strongly connected component increases rapidly \citep{barbosa03,karp90,luczak90}.  However, we are unaware of results regarding the spectra of random directed graphs.  Therefore, we numerically verified that $\lambda_1(\mathbf{A}) \sim np$ remains plausible for random directed graphs by determining the largest eigenvalue in an ensemble of 99,000 random digraphs (50 from each combination of $n = \{2, 3, \ldots,100\}$ and $p=\{0.05, 0.10,0.15,\ldots,1\})$.  Our results indicate that $\lambda_1(\mathbf{A}) \sim nC = L/n$, where $C = L/n^2$ is an estimate of $p$ (Figure~\ref{fig:sr}).  As either $L$ increases or $n$ decreases the residual error decreases.  We conclude that in random digraphs, as in undirected random graphs, \ppr is largely determined by the combination of the number of nodes and number of direct connections; pattern of connections has a minor influence.  In digraphs with a more structured topology---such as those with power-law in-degree or out-degree distributions or modularity---we might expect \ppr to deviate from $L/n$ as it does in undirected graphs, though this remains to be explored.

In ecological networks where $n$ is the number of species (functional groups, etc.) and $L$ is the number of direct transactions, the rate of pathway proliferation will be heavily influenced by species richness and direct link abundance.  However, the results of \citet{farkas01} and \citet{Goh01} suggest that if the degree distributions are skewed, as has been demonstrated for some food-webs \citep{dunne02el,montoya02,williams02}, or the networks contain other types of order such as modularity, then the residual error $d=|\lambda_1(\mathbf{A})-L/n|$ may be larger than expected from random graphs.

We have presented three key results in this section.  First, pathway proliferation occurs only if there is more than one cycle or feedback in the graph (i.e., it does not occur in acyclic networks).  Second, the dominant eigenvalue of strongly connected components $\lambda_1(\mathbf{K}_i)$ is the asymptotic proliferation rate for all nodes within that component, and this rate can vary between strongly connected components within a network.  Further, the largest $\lambda_1(\mathbf{K}_i)$ of a network is the dominant eigenvalue for the whole network such that $\max(\lambda_1(\mathbf{K}_i))=\lambda_1(\mathbf{A})$.  Third, although topology can be a factor, the proliferation rate is heavily influenced by the number of nodes and number of direct links.  Additionally, networks composed of $\alpha$ non-trivial strongly connected components $\mathbf{K}_i$ $(i = 1,\ldots, \alpha)$ have a form of structural modularity that may be functionally significant to the complex systems being modeled.

\section{Modularity and Pathway Proliferation in Food-webs}\label{sec:fw}
In this section we build on our conceptual and mathematical developments by applying the theory to determine the pathway proliferation rates in 17 of the largest empirical food-webs currently available, which are identified in Table \ref{tab:tab1}.  Five of the food-webs are terrestrial, three are from freshwater habitats, and seven represent marine or oceanic habitats.  Following accepted protocol, original food-webs were modified such that any species or functional group with identical predators and prey were grouped as a ``trophic species'' or trophospecies to reduce methodological bias in the data \citep{cohen90,pimm91, yodzis82,yodzis98}.  These food-webs have been the subject several network analyses \citep{dunne02pnas,dunne02el, dunne04,williams00,williams02} which previously reported their number of trophospecies or nodes $n$,  connectance $C = L/n^2$, the proportion of basal species $\%B$ $(k_i^{in} = 0)$, proportion of intermediate species $\%I$ $(k_i^{in},k_i^{out} >0)$ , proportion of top species $\%T$ $(k_i^{out} = 0)$, and link density $L/n$.  We reproduce this basic network information in Table \ref{tab:tab1} for comparison.  In addition, \citep{dunne02el,dunne02pnas} showed that several have skewed degree distributions (i.e., power-law, exponential).

\begin{sidewaystable}[t]
\caption{Topological properties of 17 empirical food-webs.} \label{tab:tab1}
\begin{center}
\begin{scriptsize}
\begin{tabular}{lllcccccccccccc}
\hline
\bfseries{Habitat} & \bfseries{Food Web} & \bfseries{Original Reference} & \bfseries{Taxa$^{\dagger}$} &  \bfseries{$n^{\dagger}$} &  \bfseries{$C^{\dagger}$}  & \bfseries{$\%B^{\dagger}$} & \bfseries{$\%I^{\dagger}$} & \bfseries{$\%T^{\dagger}$} & \bfseries{$L/n^{\dagger}$} & \bfseries{$\lambda_1(\mathbf{A})$}   &  \bfseries{$d$}   & \bfseries{$Pr(d)$} &   \bfseries{$\#\mathbf{K}$}  &   \bfseries{$\%\mathbf{K}$}\\
\hline
Terrestrial &   Coachella Valley    &   \citealt{polis91}  &   30  &   29  &   0.31    &   10  &   90  &   0   &   9.03 & 6.35 &  2.7 & 0.001* &  2 & 59 \\[-1 ex]
& St. Martin Island   &   \citealt{goldwasser93} &   44  &   42  &   0.12    &   14  &   69  &   17  &   4.88    &   0.00    &   4.9 &   0.001* & 0 & 0 \\[-1 ex]
&   El Verde Rainforest &   \citealt{waide96} &   156 &   155 &   0.06    &   18  &   69  &   13  &   9.74    &   10.25   &   0.5 &   0.001*   &   1   &   45 \\[-1 ex]
&   UK Grassland    &   \citealt{martinez99}    &   75  &   61  &   0.03    &   18  &   69  &   13  &   1.59    &   0.00    &   1.6 &   0.001*   &   0   &   0 \\[-1 ex]
&   Scotch Broom    &   \citealt{memmott00} &   154 &   85  &   0.03    &   1   &   40  &   59  &   2.62    &   1.00    &   1.6 &   0.001*   &   0   &   0 \\[-1 ex]
Lake/Pond   &   Skipworth Pond  &   \citealt{warren89} &   35  &   25  &   0.32    &   4   &   92  &   4   &   7.88    &   3.00    &   4.9 &   0.001*   & 2   &   20 \\[-1 ex]
&   Bridge Brook Lake   &   \citealt{havens92} &   75  &   25  &   0.17    &   32  &   68  &   0   &   4.28    &   2.00    &   2.3 &   0.001*   &   1   & 8 \\[-1 ex]
&   Little Rock Lake    &   \citealt{martinez91}   &   181 &   92  &   0.12    &   13  &   86  &   1   &   10.84   &   6.20    &   4.6 &   0.001*   &   2 & 26\\[-1 ex]
Stream  &   Canton Creek    &   \citealt{townsend98}    &   108 &   102 &   0.07    &   53  &   22  &   25  &   6.83    &   1.00    &   5.8 &   0.001* & 0 & 0 \\[-1 ex]
&   Stony Stream    &   \citealt{townsend98}    &   112 &   109 &   0.07    &   56  &   27  &   17  &   7.61    &   1.00    &   6.6 &   0.001* & 0 & 0 \\[-1 ex]
Estuary &   Chesapeake Bay  &   \citealt{baird89} &   33  &   31  &   0.07    &   16  &   52  &   32  &   2.19    &   1.00    &   1.2 & 0.001* & 0 & 0 \\[-1 ex]
&   St. Marks Estuary   &   \citealt{christian99}&   48  &   48  &   0.10    &   10  &   80  &   10  &   4.60    &   1.00    &   3.6 & 0.001* & 0 & 0 \\[-1 ex]
&  Ythan Estuary, 1991 &  \citealt{hall91} &   92  &   83  &   0.06    &   9   &   54  &   37  &   4.76    &   1.62    &   3.1 &   0.001*   &   1 & 2 \\[-1 ex]
&   Ythan Estuary, 1996 & \citealt{huxham96} &   134 &   124 &   0.04    &   4   &   56  &   40  &   4.67    &   1.62    &   3.1 &   0.001*   &   1   & 2\\[-1 ex]
Marine  &   Benguela    &   \citealt{yodzis98} &   29  &   29  &   0.24    &   7   &   93  &   0   &   7.00    &   3.00    &   4.0 &   0.001*   & 2 & 21 \\[-1 ex]
&   Caribbean Reef, small   &   \citealt{opitz96}  &   50  &   50  &   0.22    &   6   &   94  &   0   &   11.12   &   8.63    &   2.5 &   0.001* & 2 & 60 \\[-1 ex]
&   NE US Shelf &   \citealt{link02} &   81  &   79  &   0.22    &   3   &   94  &   4   &   17.76   &   4.87    &   12.9    &   0.001*   &   2 & 39 \\
\hline
\end{tabular}
\end{scriptsize}
\end{center}
\vskip 0.02in
{Taxa refers to the original number of species; $n$ is the number of nodes or trophospecies; $C =  L/n^2$ is connectance; $\%B$, $\%I$ and $\%T$ are the proportions of basal (indegree~$= 0$), intermediate (indegree and outdegree~$> 0$), and top (outdegree~$= 0$) trophospecies; $L/n$ is link density; \ppr is the dominant eigenvalue of the entire foodweb; $d = |\lambda_1(\mathbf{A}) - L/n|$; $Pr(d)$ is the fraction of an ensemble of 1001 random digraphs in which $d$ is greater than or equal to that observed in $\mathbf{K}_i$; * indicates statistically significant $Pr(d)$ at $\alpha = 0.05$; $\#\mathbf{K}$ is the number of non-trivial strongly connected components; $\%\mathbf{K}$ is the percent of species in a non-trivial strongly connected component.  $\dagger$ marks topological properties previously reported for these food-webs \citep{dunne02pnas,dunne04,williams00}.}
\end{sidewaystable}

\subsection{Methods}
We first identified and characterized all strongly connected components ($\mathbf{K}_i$), including their rates of pathway proliferation and damping ratio, in these food-webs.  We envisioned three possible outcomes.  If food-webs were adequately modeled by random digraphs then we would expect each web to have one strongly connected component encompassing most if not all of the nodes with a single pathway proliferation rate close to link density  $\lambda_1(\mathbf{A}) \sim L/n$.  This seemed unlikely given the known skewed degree distributions and additional evidence that ecological processes construct non-random topologies \citep{cohen90,williams00}, despite arguments to the contrary \citep{kenny91}.  A second possibility is based on the observation that most early food-webs were acyclic \citep{cohen90}.  Thus, the 17 food-webs in our study, all developed since 1990, might also be acyclic digraphs with no non-trivial strongly connected components.  In this case, pathway proliferation would not occur and $\lambda_1(\mathbf{A}) = 0$.  A final possibility is that the food-webs would tend to have one or more strongly connected components and multiple pathway proliferation rates.  This outcome would support the hypothesized modularity of ecological systems which is thought to increase system stability \citep{krause03,may72,pimm80,yodzis82}.

In Section \ref{sec:pp} we hypothesized that the absolute difference between the dominant eigenvalue and its expected value in random digraphs ($L/n$) might be a useful indicator of the significance of network topology.  To assess this hypothesis, we used Monte Carlo simulations to determine if $d=|\lambda_1(\mathbf{A})-L/n|$ was larger than expected.  We had two scales of analysis: whole food-web and non-trivial strongly connected components.  For both, we constructed 1001 uniform random digraphs with $n$ nodes, where each possible link was connected with probability $p$ equal to the original network's connectivity ($p = C$).  We assessed statistical significance by determining the fraction of random digraphs in which $d$ was equal or greater than observed in our network of interest, $Pr(d)$.  Assuming a significance level of $\alpha = 0.05$, $Pr(d) < 0.05$ implies $d$ is statistically significant.

When applied to the entire food-web, a significant difference with the null model implies that topological factors beyond species and link richness are significant in determining the whole system dominant eigenvalue.  This could be the size or frequency of strongly connected components within the network, as suggested by the analysis in Section \ref{sec:pp}, or perhaps a skewed degree distribution.  If the food-webs had more than one non-trivial component, then we expected the deviation to be large.

When applied to strongly connected components, a significant deviation of $d$ also indicates the importance of network topology.  However, given that they are irreducible by definition, a significant deviation of $d$ within a component must indicate the significance of another element of topology, such as the degree distribution.
\subsection{Results}
Food-webs included in this study range in size from 25 to 155 trophospecies and 3$\%$ to 32$\%$ connectance (Table \ref{tab:tab1}).  They tend to have a large proportion of intermediate species (i.e., those with $k_j^{in}>0$  and $k_j^{out}>0$), although the two stream food-webs are notable exceptions.  Ten of the 17 food-webs examined contained at least one non-trivial strongly connected component; six had two.  Five of the remaining food-webs had a dominant eigenvalue of unity, implying that at least one node contained a self-loop.  Our results reveal that the majority of these food-webs have at least one directed cycle, contrary to earlier food-web theory \citep{cohen90}.

While the majority of the food-webs have a modular organization that is based on strongly connected components, the proportion of species involved in the modules is variable.  In food-webs that have a non-trivial strongly connected component, the proportion of the original nodes involved ranges from 2$\%$ in the two Ythan Estuary food-webs to 60$\%$ in the Caribbean reef model.  Notice that the definition of a strongly connected component bans nodes that have no inputs or no outputs, which excludes basal species $(k_j^{in}=0)$ and top consumer species $(k_j^{out}=0)$. Therefore, the total number of species in strongly connected components is limited by the number of intermediate species.  This may be a factor in why the two stream food-webs and the Scotch Broom food-web contain no non-trivial components.

The absolute difference between the dominant eigenvalue of the entire food-web and its expected value based on random digraphs of the appropriate size and connectance $(d=|\lambda_1(\mathbf{A})-L/n|)$ ranged from a minimum of 0.5 for the El Verde rainforest to a maximum of 12.9 for the NE US shelf food-web (Table \ref{tab:tab1}).  In all cases, this difference was significantly different from the random digraph null model, indicating that topology is a significant factor in determining \ppr.  This result is consistent with the presence of one or more small non-trivial components and acyclic digraphs.  Our results provide another line of evidence suggesting that the ecological processes that create food-webs lead to more ordered network topologies; random digraphs are not good models for these systems.

Inspection of the strongly connected components reveals a diversity of topologies as shown in Table \ref{tab:tab2}.  The largest strongly connected component, with seventy trophospecies occurs in the El Verde rainforest model; although it is the least well connected (13$\%$), it still has the largest rate of pathway proliferation $(\lambda_1(\mathbf{A}) = 10.25)$.  In contrast, ten of the sixteen strongly connected components have four or fewer species.  Five of the strongly connected components only contain two trophospecies, requiring a single simple cycle of path length 2 (e.g., $j \rightarrow i \rightarrow j$).  While the two strongly connected components in the Coachella Valley, Skipworth pond, and Benguela are about the same size, one of the two components in Little Rock Lake, Caribbean reef, and NE US shelf is substantially larger than the other.  Table \ref{tab:tab3} lists the trophospecies in the two strongly connected components of the Coachella Valley.

\begin{sidewaystable}[t]
\caption{Topological properties of strongly connected components in 17 empirical food-webs} \label{tab:tab2}
\begin{small}
\begin{center}
\begin{tabular}{lcccccccccc}
\hline
\textbf{Model}& $\mathbf{K}_i$ & $n$ &  $L$ &  $C$ &  $L/n$& $\lambda_1(\mathbf{K}_i)$ & mult($\lambda_1$)& $\rho$ & $d$ &$Pr(d)$\\
\hline
Coachella Valley & 1 & 11 & 71 & 0.59 & 6.45 & 6.35 & 1 & 3.40 & 0.1 & 0.001*\\[-1 ex]
& 2 & 6 & 22 & 0.61 & 3.67 & 3.56 & 1 & 3.56 & 0.1 & 0.001*\\[-1 ex]
El Verde Rainforest & 1 & 70 & 633 & 0.13 & 9.04 & 10.25 & 1 & 2.43 & 1.2 & 0.001*\\[-1 ex]
Skipworth Pond & 1 & 3 & 9 & 1.00 & 3.00 & 3.00 & 1 & -- & 0.0 & 1.001\\[-1 ex]
 & 2 & 2 & 4 & 1.00 & 2.00 & 2.00 & 1 & -- & 0.0 & 1.001\\[-1 ex]
Bridge Brook Lake & 1 & 2 & 4 & 1.00 & 2.00 & 2.00 & 1 & -- & 0.0 & 1.001\\[-1 ex]
Little Rock Lake & 1 & 21 & 167 & 0.38 & 7.95 & 6.20 & 1 & 2.42 & 1.7 & 0.001*\\[-1 ex]
  & 2 & 3 & 9 & 1.00 & 3.00 & 3.00 & 1 & -- & 0.0 & 1.001 \\[-1 ex]
Ythan Estuary, 1991 & 1 & 2 & 3 & 0.75 & 1.50 & 1.62 & 1 & 2.62 & 0.1 & 0.212\\[-1 ex]
Ythan Estuary, 1996 & 1 & 2 & 3 & 0.75 & 1.50 & 1.62 & 1 & 2.62 & 0.1 & 0.226\\[-1 ex]
Benguela  & 1 & 3 & 9  &   1.00 &     3.00  &    3.00 &     1 &  -- &  0.0 & 1.001\\[-1 ex]
 &  2  &  3  & 7 & 0.78 &  2.33 & 2.25  &  1 &  4.05 & 0.1 & 0.007*\\[-1 ex]
Caribbean Reef, small & 1  & 2 & 3 & 0.75 & 1.50 & 1.62 & 1 & 2.62 & 0.1 & 0.217\\[-1 ex]
 & 2 & 28 & 244 & 0.31 & 8.71 & 8.63 & 1 &  4.11 &  0.1 & 0.001*\\[-1 ex]
NE US Shelf & 1 & 4 & 11 & 0.69 & 2.75 & 2.88 & 1 & 4.41 & 0.1 & 0.003*\\[-1 ex]
 & 2 & 27 & 243 & 0.33 & 9.00 & 4.87 & 1 & 1.69 & 4.1 & 0.001*\\
\hline
\end{tabular}
\end{center}
\vskip 0.02in
{$\mathbf{K}_i$ indicates the non-trivial strongly connected component number; $n$ is the number of nodes (trophospecies), $L$ is the number of links; $C = L/n^2$ is connectance, $L/n$ is the link density, $\lambda_1(\mathbf{K}_i)$ is the dominant eigenvalue of $\mathbf{K}_i$, mult($\lambda_1$) is the multiplicity of the dominant eigenvalue, $\rho=\lambda_1(\mathbf{K}_i)/\|\lambda_2(\mathbf{K}_i)\|$ is the damping ratio (-- indicates $\rho$ is undefined because $\|\lambda_2(\mathbf{K}_i)\|=0$), $d=|\lambda_1(\mathbf{A})-L/n|$ is the absolute difference between the dominant eigenvalue and link density, $Pr(d)$ is the fraction of an ensemble of 1001 random digraphs in which $d$ is greater than or equal to that observed in $\mathbf{K}_i$, and $*$ indicates statistically significant $Pr(d)$ at $\alpha = 0.05$.\vskip 0.01in}
\end{small}
\end{sidewaystable}


\begin{table}[t]
\caption{Trophospecies in the two non-trivial strongly connected components of the Coachella Valley food-web.} \label{tab:tab3}
\begin{small}
\tableline\vskip -0.5ex
\begin{tabular}{ll}
\bfseries{$\mathbf{K}_1$} & \bfseries{$\mathbf{K}_2$}\\[0 ex]
\hline
primarily herbivorous mammals and birds &small arthropod predators\\[-1 ex]
small omnivorous mammals and birds & medium arthropod predators\\[-1 ex]
primarily carnivorous lizards & large arthropod predators\\[-1 ex]
primarily carnivorous snakes & facultative arthropod predators\\[-1 ex]
large primarily predacious birds & life-history arthropod omnivore\\[-1 ex]
large primarily predacious mammals & spider parasitoids\\[-1 ex]
& primary parasitoids\\[-1 ex]
& hyperparisitoids\\[-1 ex]
& predacious mammals and birds\\[-1 ex]
& arthropodivorous snakes\\[-1 ex]
& primarily arthropodivorous lizards
\end{tabular}
\tableline
\end{small}
\end{table}

The dominant eigenvalues of all strongly connected components have a multiplicity of one, so the adjacency matrices associated with the component subdigraphs are primitive.  Therefore, the dominant eigenvalues represent the strongly connected component asymptotic rates of pathway proliferation.  These range from 1.62 in strongly connected components of the two Ythan Estuary food-webs to 10.25 in the large El Verde rainforest strongly connected component and generally increase with link density as would be expected in random digraphs.  However, half of the strongly connected components have a statistically significant difference between the dominant eigenvalue and link density including Coachella Valley ($\mathbf{K}_1$ and $\mathbf{K}_2$), El Verde Rainforest ($\mathbf{K}_1$), Little Rock Lake ($\mathbf{K}_1$), Benguela ($\mathbf{K}_2$), Caribbean Reef ($\mathbf{K}_2$), and NE US shelf ($\mathbf{K}_1$ and $\mathbf{K}_2$).  The topological arrangement of species and links in these three strongly connected components influences their rate of pathway proliferation; the others are largely determined by their species and link richness.

The damping ratio defined in equation \ref{eq:damp} is an index of the speed of convergence to the asymptotic rate of pathway proliferation; A larger ratio indicates faster convergence.  Five of the strongly connected components are completely connected.  They have a pathway proliferation rate equal to their trophospecies richness and an undefined damping ratio because their second eigenvalues are zero.  In these cases the asymptotic rate of pathway proliferation is achieved instantaneously.  The other damping ratios range from 1.69 in $\mathbf{K}_2$ of the NE US Shelf to 4.41 in $\mathbf{K}_1$ of the same food-web.  Transient dynamics of the pathway proliferation rate, determined by the smaller eigenvalues, are more influential in NE US shelf ($\mathbf{K}_2$).  Its pathway proliferation rate does not converge until a pathway length of nearly twenty-two, while in NE US shelf ($\mathbf{K}_1$) the proliferation rate converges by a pathway length of eight.

In summary, the majority of the food-webs we examined contained at least one non-trivial strongly connected component.  Six food-webs had two non-trivial strongly connected components; none had more than two.  The proportion of species involved in a strongly connected component ranged from 2$\%$ to 60$\%$.  In all cases, the difference between the dominant eigenvalue of the food-web and the expected value ($L/n$) in a random network was significant.  This difference occurs because the topology of food-webs is non-random; thermodynamic processes and species characteristics (e.g., the species niche) combine to form non-random structures \citep{chase03}.  Within the strongly connected components, the rate of pathway proliferation ranged from 1.62 to 10.25 and half were indistinguishable from random graphs based on the expected rate of pathway proliferation.
\section{Discussion}
As with any analysis of network models that reveals previously hidden structural patterns, we are left with two questions.  First, what, if any, significance do these patterns hold for our systems of interest?  Are strongly connected components and pathway proliferation simply another pretty pattern, another network or food-web statistic to report, or do they impart some functional significance?  The second question cannot be divorced from the first; what system processes might create these structural patterns?  Are there ecological processes or forces that might lead to the development of these structures?  These are not easy questions to answer, but in this section we attempt to address them for the presence of strongly connected components and pathway proliferation in ecological networks.

Strongly connected components introduce a form of modularity into network models, where modularity is defined as a hierarchical system subdivision into more or less interacting subsystems.  Several types of modularity have been proposed in ecological systems.  Building on earlier ideas in general systems theory that linked hierarchical organization to system stability \citep[e.g.][]{simon62}, \citet{may73} hypothesized that ecosystems have modular structures.  He found that the Lyapunov stability of randomly assembled ecosystems tended to be greater when the species were partitioned into blocks of interacting species with few if any connections to other blocks.  \citet{pimm79} termed these blocks of species ``compartments'', stating that they are ``...characterized by strong interactions within compartments, but weak interactions among compartments'' (p. 145).  \citet{pimm80} concluded from a study of binary empirical food-webs that, while there was evidence species were grouped into subsystems largely by habitat, compartmentalization as defined by Pimm was an uncommon phenomenon. They noted, however, that a complete test of the hypothesis would require knowledge of the strength of interactions, which was absent in their food-web models.  It is also possible that they were unable to identify compartmentalization in their food-webs because their models exclude detrital recycling.  However, \citet{neutel02} showed that even without detrital recycling, long, weak links enhanced system stability.

Yodzis (1982) remarked that modular organization was an old idea in ecology, citing the guild concept \citep{root67} as an example.  He applied the dominant clique idea from graph theory to identify another type of modularity in food-webs.  He defined ``a clique as a set of species in a given ecosystem with the property that every pair in the set has some food resource in common, and $\ldots$ a dominant clique as a clique which is contained in no other clique'' (p. 552).  More recently, \citet{krause03} used a methodology developed to identify cohesive subgroups in social network analysis to classify another type of modularity in food-webs.  They demonstrated that this type of organization increased system stability to species deletion by localizing the effect within a module.  In addition, \citet{allesina2005} found four or more modules based on strongly connected components in 17 network models of carbon flux. The four modules usually grouped into four types: inputs, outputs, dissipation, and species and nutrient pools.  Their work is the most similar to that presented in this paper, but there are two important differences.  First, we examined different types of ecological networks.  They examined ecosystem flow networks, while we restricted our analysis to food-webs.  Second, we considered inputs, outputs, and dissipation to be external to the system and therefore we could not identify these as separate modules.

The dominant ecological hypothesis is that food-web or ecosystem modularity increases overall system stability by localizing interactions within modules.  Given the static, binary, presence-absence information of food-webs in our study we were unable to meaningfully test this hypothesis; stability is inherently a dynamic concept.  Known issues with food-web model construction further make this hypothesis difficult to resolve \citep{cohen93,polis96}.  Empirical food-web models usually do not indicate interaction strength or the temporal and spatial variation of the interactions, as these details are expensive to acquire.

Despite the challenge of assessing their effect on system stability, strongly connected components in ecological networks appear to be important functional elements of system organization and provide new insights about species participating in them.  By definition (Section \ref{sec:pp}), there is minimally one simple cycle that encircles all nodes in the strongly connected component.  This provides at least one channel for cybernetic feedback (positive or negative) to propagate among species in the module \citep{deangelis86, patten59,patten81}.  Furthermore, in food-webs it is reasonable to assume that each predatory species directly benefits by its consumption of prey.  This establishes an indirect mutualism that spans the strongly connected component, and provides the necessary conditions for the strongly connected components of food-webs to function as autocatalytic cycles---systems that catalyze their own production \citep{smith95,ulanowicz97}.   Autocatalytic cycles are an essential element of metabolism in chemical and living systems and may have played a role in the origin of life \citep{smith95}.  \citet{smith95} describe autocatalytic cycles as a force for cooperation among the member species and efficient information integrators.

\citet{ulanowicz97} identifies several emergent properties autocatalytic cycles may possess, including centripetality, persistence, and autonomy.  Centripetality is the tendency of the cycle to attract more of the energy--matter flux of the system. If any member species becomes more efficient at using its resources or better able to acquire new resources such that its population increases, this positive change tends to cascade through the module, collectively benefiting the populations of all species involved.  Autocatalytic cycles tend to persist because their general form can be maintained in a system despite fluctuations in the interaction strengths and possible element replacement.  In food-webs this implies that when a trophically similar species is introduced to the system, if it is more efficient or in some way ecologically more competitive, it may wholly replace an existing species in the autocatalytic cycle, but the cycle remains.  Finally, autocatalytic cycles can establish a degree of autonomy because species in the cycle can actively influence at least a portion of their input environment.  In this sense, species in strongly connected components of food-webs are involved in ecosystem engineering \citep{jones97} and niche construction \citep{laland99,oldingsmee03}.  \citet{ulanowicz86, ulanowicz97} further argues that the autocatalytic nature of cycles in ecosystems makes them a principal agent in ecosystem growth and development.

\citet{smith95} remarked that autocatalytic cycles are sensitive to cheaters or parasites that feed off the strongly connected component without participating in the cycle.  Top predators feeding on species in a strongly connected component or a downstream strongly connected component might function as parasites in this sense.  Perhaps this is why strongly connected components do not occur in all the food-webs analyzed.  This and the tendency for centripetality may explain why there are fewer than three strongly connected components in these food webs.  However, we are unable to assess these possibilities with these data because the differences may simply reflect disparities in food-web modeling decisions.

Pathway proliferation rates of strongly connected components provide additional information about these modules.  Each additional link in a strongly connected component beyond those that form the defining cycle introduces another embedded simple cycle.  This lowers the maximal distance between nodes in the module, increases the potential pathways for energy--matter flux, tends to increase the rate of pathway proliferation, and leads to the unbounded growth of pathways as length increases.  In some cases, the rate of pathway proliferation will not increase as expected with the number of links.  For example, half of the strongly connected components identified in our food-webs had pathway proliferation rates that were significantly different from the expected rate based on random graphs with a Poisson degree distribution.  This suggests that module topology differs from what we would expect from a random generating process.

As mentioned earlier, we expected the ecological processes forming food-webs to generate non-random structures.  Species characteristics such as metabolic requirements, food preferences, capture ability and handling time, as well as other niche requirements and natural history constraints should directly influence the choices of ``who eats whom'' and how much.  In addition, the emergent properties of autocatalytic cycles and ecosystems more generally may provide whole-system constraints.

Perhaps the interesting question is not, why the eight strongly connected components did not match the random expectation, but why the other half did?   Notice that the strongly connected components with non-random topologies were the largest modules, while the eight strongly connected components with topologies indistinguishable from random digraphs involved only two or three trophospecies.  The universe of possible topologies is much smaller in these small and well-connected ($0.75 \le C \le 1$) modules, making the ecologically created topologies reflected in the food-webs more likely.  Five of these modules were completely connected, generating only one possible arrangement.  The eight strongly connected components with apparently non-random topologies were less well connected ($0.13 \le C \le 0.78$), generating a much larger universe of possible topologies.  In six cases the pathway proliferation rate was significantly less than expected, but in the large strongly connected component of the El Verde rainforest and the smaller module of the Caribbean Reef, the pathway proliferation rate was more than expected.  At this point, we are unable to provide a satisfactory explanation for these differences.

In our discussion thus far, we have been interchanging food-webs and ecosystems.  It is important to recognize, however, that food-webs are a subset of a broader class of ecosystem models of energy--matter flow.  Food-webs are generally defined by the relation ``who eats whom'' that is one process generating energy--matter flux, while ecosystem flow--storage models typically trace a conserved flow unit (e.g., energy, nitrogen, phosphorus) through the system, regardless of the process producing the flow.  Thus, non-trophic ecological processes such as excretion and death are captured in flow-storage models, revealing a different picture of ecosystem organization with additional cycling.

Pathway proliferation influences the development and significance of indirect flows in ecosystem flow--storage models.  Indirect flows are an important aspect of the ecological significance of the strongly connected components, so here we take a closer look.  Indirect flows are derived from two fundamentally distinct types of pathways: chains (e.g., $5 \rightarrow 6 \rightarrow 7$ in Figure~\ref{fig1}a) and cycles (e.g., $1 \rightarrow 3 \rightarrow 1$ in Figure~ \ref{fig1}a).  Indirect flows in chains are limited by transfer efficiencies and chain length.  In cycles, the number and length of pathways are unlimited such that indirect flows are only limited by transfer efficiencies reflecting energy--matter dissipation and export.  As ecosystems are open thermodynamic systems, shorter indirect pathways individually will tend to carry larger indirect flows than longer indirect pathways.  A faster rate of pathway proliferation \ppr implies that there will be more shorter indirect pathways, increasing the possibility that the magnitude of indirect flows will surpass that of direct flows.  Thus, within a strongly connected component the large number of indirect pathways will tend to carry a large fraction of the flow between species (nodes).

More generally, \ppr indicates the potential for direct and indirect energy, matter, and information transmission between compartments in a strongly connected component.  Realized transmission rates are dependent on the realized use of each pathway.  Previous ecosystem network analyses reveal some of the system-level consequences of differential pathway use \citep{fath04, fath98,fath99, higashi89, patten85b, ulanowicz86}, but there is much left to learn about this subject.  The interplay of this potential and realized network structure is an interesting, important topic for understanding the organization and transformation of complex adaptive systems like ecosystems.

We conclude that the strongly connected components and pathway proliferation are ecologically relevant phenomena because they provide novel insights about the system of interest.  Without knowing the strength of interactions or energy--matter flux rates, the presence of these structural features suggests groups of species functionally integrated by indirect effects mediated by autocatalytic cycles.  They portend the possibility of integral species relationships that are shifted toward more positive associations and the possibility of the dominance of indirect flows.  In some cases, apparent negative interactions such as predation or competition may become more positive through indirect interactions mediated by the autocatalytic cycles of the strongly connected components.
\section{Acknowledgements}
We are grateful to Jennifer Dunne for providing the food-web data, and discussions with Stuart Whipple and the University of Georgia Systems and Engineering Ecology group that enhanced this work.  The manuscript benefited from reviews by M.B. Beck, H.R. Pulliam, P.G. Verity, D.K. Gattie, S. Bata, and two anonymous peers.  SRB was supported by National Science Foundation Grants No. OPP-00-83381 and IIS-0326059.

\end{spacing}

\newpage
\clearpage

\end{document}